\RequirePackage[l2tabu, orthodox]{nag}
\documentclass{llncs}
\usepackage[utf8]{inputenc}
\usepackage{graphicx}
\usepackage{amsmath}
\usepackage{amssymb}
\usepackage{mathtools}
\usepackage{multirow}
\usepackage{stmaryrd}
\usepackage{multirow}
\usepackage{etoolbox}
\usepackage{moresize}
\usepackage{placeins}
\usepackage[inline]{enumitem}

\usepackage[hyphens]{url}
\usepackage[breaklinks]{hyperref}
\usepackage{booktabs} 
\usepackage[bottom]{footmisc}
\usepackage[ruled,linesnumbered]{algorithm2e}
  \SetAlFnt{\scriptsize\sf}

\usepackage{subcaption}
\usepackage{todonotes}

\def\ptFiguresDirectory#1{./figures/#1}

\def\FWER#1{FWER}

\newcommand{\stiny}{\fontsize{4pt}{5pt}\selectfont}

\begin{document}
\frontmatter          
\pagestyle{headings}  
\addtocmark{} 
\mainmatter              
\title{Cascade of one-class classifier ensemble and dynamic naive Bayes classifier applied to the myoelectric-based upper limb prosthesis control with contaminated channels detection}
\titlerunning{Cascade of one-class classifier ensemble and dynamic naive Bayes classifier}  
%
\author{Pawel Trajdos\orcidID{0000-0002-4337-6847} \and Marek Kurzynski\orcidID{0000-0002-0401-2725}}
\authorrunning{P. Trajdos \and M. Kurzynski} 
%
\tocauthor{Pawel Trajdos, Marek Kurzynski}
\institute{Wroclaw University of Science and Technology, Wroclaw, Poland,\\
\email{pawel.trajdos@pwr.edu.pl}, \email{marek.kurzynski@pwr.edu.pl}}

\maketitle              

\begin{abstract}

Modern upper limb bioprostheses are typically controlled by sEMG signals
using a pattern recognition scheme in the control process. Unfortunately, the sEMG signal is very susceptible to contamination that deteriorates the quality of the control system and reduces the usefulness of the prosthesis in the patient's everyday life. 
In the paper, the authors propose a new recognition system intended for sEMG-based control of the hand prosthesis with detection of contaminated sEMG signals. 
The originality of the proposed solution lies in the co-operation of two recognition systems working in a cascade structure: (1) an ensemble of one-class classifiers used to recognise contaminated signals and (2) a naive Bayes classifier (NBC) which recognises the patient's intentions using the information about contaminations produced by the ensemble. 
Although in the proposed approach, the NBC model is changed dynamically, due to the multiplicative form of the classification functions, training can be performed in a one-shot procedure.
Experimental studies were conducted using real sEMG signals. The results obtained confirm the hypothesis that the use of the one-class classifier ensemble and the dynamic NBC model leads to improved classification quality.

\end{abstract}

\section{Introduction} \label{sec:Introduction}


In the 1950s, the concept of controlling an active, externally powered hand prosthesis using an sEMG signal was introduced. This initiated the intensive development of advanced myoelectrically controlled anthropomorphic upper limb prostheses, which is still ongoing \cite{Chen2023}. 
Modern bioprostheses are typically controlled by sEMG signals from residual limb muscles
using a pattern recognition scheme in the control process. 
In this approach, the patient's intention of moving the prosthesis (manipulation or grip)  denotes the object being recognised, the recorded sEMG signals constitute the object representation, and the type of movement of the prosthesis is the class label \cite{Trajdos2024}.



This means that the correct operation of the control system, which determines the comfort of using the prosthesis in patient's everyday life, depends exclusively on the quality of the overall system for recognizing the patient's intention.
The essence of the recognition system is a sequence of actions performed on the electromyographic signal by a human (patient), a measuring system, and a classifier. Each stage of this sequence may be a source of errors that deteriorate the quality of the control system \cite{Farago2023}.


The topic of this paper is to address the recognition of contaminated sEMG signals within the context of myoelectric control for upper limb prosthetics in order to improve the quality of the control process.

Due to its low power and unfavourable recording conditions, the sEMG signal is very susceptible to contaminations, which can be divided into three main categories \cite{Kyranou2018}:
\begin{enumerate*}
    \item  noise (thermal and flicker noise, amplifier saturation, analogue to digital signal clipping, quantisation noise),

    \item external interference (power line, ECG, crosstalk),

    \item artifacts (measurement artifacts, baseline wander, motion artifact, electrode shifts, changes of electrode/skin impedance (sweating))
    
\end{enumerate*}

Generally, two types of methods are proposed in the literature to mitigate the adverse effects of sEMG signal contamination.
The first type of methods is aimed at directly improving the quality of the recorded sEMG signal, which can be achieved by better hardware \cite{Farago2023B} or by software filters tailored to characteristics of different contamination of  sEMG signal \cite{Boyer2023}. 

The second category includes methods based on the artificial intelligence paradigm, in which information about sEMG signal contamination is extracted, and then this information is used in model of a decision-making system to improve the quality of the control process.  

Particularly noteworthy are the methods with a common concept based on the use of two multiclassifier recognition systems. The first system is an ensemble of one-class classifiers that assess the degree of contamination of individual sEMG signals (channels).
The purpose of the second system is to classify the movement of the prosthesis, and its recognition procedure is supported by the results from a one-class classifier ensemble. 
 The advantage of this solution is the completeness of the patient's intention recognition with detection of contaminated channels and independence of the type of contamination, because one-class classifiers are trained exclusively on clean signals. This concept is quite richly represented in the literature \cite{Fraser2014}, \cite{Furukawa2015}, including the authors' earlier works \cite{Trajdos24b}, \cite{Trajdos2025} . The proposed solutions differ in the decision scheme of the one-class classifiers (the method of assessing signal contamination), the number of channels (signals) used in the base classifiers of the second system and the classifiers combining procedure.

In the paper, the authors develop a new recognition system intended for sEMG-based control of the upper limb prosthesis with detection of contaminated sEMG signals, which also fits into the concept described above. 
The originality of the proposed solution lies in the co-operation of an ensemble of single-class classifiers and a naive Bayes classifier working in a cascade structure. 
In the first step, the ensemble of one-class classifiers is typically used to recognise contaminated channels, returning information about the level of contamination for each channel. In the second step, the naive Bayes classifier recognises the patient's intentions using the information produced by the ensemble. This means that the model of the naive Bayes classifier is dynamically changing, as the information about contamination may differ between objects. The choice of the naive Bayes classifier for recognition of patient's intention was deliberate and justified by the multiplicative form of its classification functions, which allowed for training the classifier in a one-shot procedure.

The main goal of the paper and the conducted experimental study is to answer the following research questions:
\begin{itemize*}
	
	\item \label{itm:rq:1} \textbf{RQ1} Does the proposed method perform better than a simple classifier trained on the entire feature space?

	\item \label{itm:rq:2} \textbf{RQ2} Does the proposed method perform better than the alternative approaches presented in the literature?

	\item \label{itm:rq:3} \textbf{RQ3} How do the investigated methods perform under different values of signal-to-noise ratio?

\end{itemize*}

The paper is organised as follows. Section~\ref{sec:Methods} presents the models of two classification systems in a cascade structure and the method for their cooperation.  Section~\ref{sec:ExpSetup} describes the experimental setup. The results of experimental studies with discussion are presented in Section~\ref{sec:ResAndDisc}. Section~\ref{sec:Conclusions} concludes the paper.

\section {Methods}\label{sec:Methods}

Let us consider a bioprosthetic hand control system based on a multi-channel sEMG signal recognition scheme. 
Let
\begin{equation}  \label{1}
\mathcal{C}=\{C_1, C_2, \ldots, C_L\}
\end{equation}
denotes the set of sEMG signals recorded from the $L$ sensors (channels) located on the patient's forearm stump. The signals \eqref{1} represent the patient's intention to perform a specific movement of the prosthesis. 
The activation of residual stump muscles, which are the source of sEMG signals, is based on the phantom movement of the amputated limb that the user is able to voluntarily control \cite{Akbulut2022}.  

Let $x_l=(x_l^{(1)}, x_l^{(2)}, \ldots, x_l^{(d_l)})$ be a feature vector belonging to the feature space $\mathcal{X}_l$, which was extracted from the sEMG signal $C_l$ ($l=1,2, \ldots,L$). Consequently,  $x=(x_1, x_2, \ldots, x_L) \in \mathcal{X}$ denotes the whole feature vector describing the patient's intention, and 
 $\mathcal{X}=\mathcal{X}_1 \times \mathcal{X}_2 \times \ldots \times \mathcal{X}_L$.

Movements executed by the prosthesis are labeled by consecutive natural numbers, forming a set of classes for the recognition task: 
\begin{equation}   \label{2}
\mathcal{M}=\{1,2, \ldots , M\}.
\end{equation}
The number of classes $M$ can vary among patients, as it is related to their ability to activate the muscles of the stump. 

We assume that sEMG signals \eqref{1} can be contaminated during the acquisition process. Contaminants can randomly occur in various sEMG signal channels, making it impossible to predict in advance the number of affected channels and their identities.

Signal contaminants can considerably reduce the classification quality of a recognition system. Therefore, our objective is to develop a recognition system capable of identifying signal contamination and leveraging these data to enhance the classification quality of contaminated samples. 

The proposed method is based on two cooperating classification systems in a cascade structure. The first system is an ensemble of one-class classifiers whose task is to recognise contaminated sEMG signals.  

The second system is a Naive Bayes classifier (NB) working in dynamic mode and using the information provided by the previous one-class ensemble system.

The following subsections elaborate on the specifics of both classifier systems.

\subsection{Ensemble of one-class classifiers}\label{sec:Methods:oneclass}

An ensemble of one-class classifiers is denoted as:.
\begin{equation}   \label{3}
    \Phi=\left\{ \phi_1, \phi_2, \ldots, \phi_L \right\}.
\end{equation}
The base classifier $\phi_l$ of the system \eqref{3} uses 
the feature vector  $x_l$ extracted from signal $C_l$ and produces a prediction
\begin{equation}    \label{4}
\phi_l(x_l) = r_l, \;  l=1,2, \ldots ,L,
\end{equation}
that evaluates the contamination of signal $C_l$. The interpretation of the prediction \eqref{4} depends on the decision scheme adopted by the system \eqref{3}.

\textbf{Crisp decision scheme}, in which $r_l \in \{0, 1\}$.
Zero indicates that $x_l$ is an outlier in the feature space $\mathcal{X}_l$ and the corresponding signal $C_l$ is recognised as contaminated with noise. Consequently, $r_l=1$ indicates that $x_l$ is recognised as a target class object, i.e., the corresponding signal $C_l$ is considered free of contaminants. 

\textbf{Soft decision scheme}, in which $r_l \in [0, 1]$.
 The closer to 0 $r_l$, the level of contaminants in the registered signal is greater.

Moreover, let $r=(r_1, r_2, \ldots , r_L) \in \mathcal{R}$ be a vector of predictions produced by the ensemble $\Phi$, where $\mathcal{R} = \{0, 1\}^L$ ($= [0, 1]^L$) for the crisp (soft) scheme, respectively. 
 
Each base classifier \eqref{4} is trained using a channel-specific training set that contains only objects of the target class without the presence of outliers (contaminated signals). This is the usual way to build one-class classifiers~\cite{Seliya2021}. 

\subsection{Naive Bayes classifier in dynamic mode}\label{sec:ProposedMethod:multiclass}

The second classifier in the cascade is designed to recognise the class of prosthesis movement from the set \eqref{2} according to the patient's intention:

\begin{equation}   \label{5}
\psi(x, r) :  \; \;  \mathcal{X} \times \mathcal{R}  \; \rightarrow \; \mathcal{M}.
\end{equation}

The classifier \eqref{5}, in addition to the typical dependence on the feature vector $x$, uses additional information on the contamination of individual sEMG signals $C_l$ contained in the prediction vector $r$ generated by the one-class system \eqref{3}.  

Now the key question arises: How can the prediction vector $r$ in the classifier \eqref{5} be used to achieve improvement in classification quality?
In the proposed method, this is done by eliminating features originating from contaminated signals (for the crisp scheme) or by weighting the influence of features depending on the degree of channel contamination (for the soft scheme).

 To demonstrate this mechanism in detail, let us assume the canonical classifier model
which means that for a given feature vector $x$ and for prediction vector $r$ the classifier produces a vector of supports  (values of the classifying functions) for each class:
\begin{equation}   \label{6}
d(x,r)=[d_1(x,r), d_2(x,r), \ldots, d_M(x,r)],
\end{equation}
where $d_j(x,r) \in [0, 1]$ and $\sum_j d_j(x,r)=1$. The class with the highest support is the final recognition result of the classifier  $\psi$. 

For this classification model, the mechanism of influence of the prediction vector $r$ on classifier performance is quite intuitive. Namely, it is necessary to increase (decrease) the value of supports depending on the purity (contamination) of the signals from which the features are extracted.
Unfortunately, this approach has a significant drawback, as the signal contamination phenomenon is dynamic in nature, which means that the configuration of contaminated channels will probably be different for each recognised object. This requires dynamic (for each object) training of the classifier $\psi$.
Such a solution is out of the question because the bioprosthesis control system is a real-time system (the time from the beginning of sEMG signal recording to prosthesis movement should not be longer than 200 ms).

A classifier with a Naive Bayes (NB) model can be helpful here, as it eliminates the problem of dynamic learning due to the assumption of feature independence. The NB classifier is based on a probabilistic model in which the supports are equal to the posterior probabilities for the individual classes $d_j(x,r)=p(j|x;r)$, $j \in \mathcal{M}$, where $r$ denotes the parameter of these probabilities. Using Bayes' formula we can express the posterior probabilities in the form:
\begin{equation}  \label{7}
p(j|x;r)=c \cdot p_j \cdot P(x|j;r), \  \  j \in \mathcal{M},
\end{equation}
where $c$ denotes normalizing factor, $p_j$ is a prior probability of the $j$-th class and $P(\cdot )$ denotes probability distribution of feature vector $x$ given that object is from the $j$th class.

Assuming conditional independence of features, we have
\begin{equation}  \label{8}
P(x|j;r) = \prod_{l=1}^{L} \prod_{i=1}^{d_l} P(x_l^{(i)}|j;r)
\end{equation}
where $P(\cdot )$ denotes the conditional probability distribution (probability or density function for a discrete or continuous feature, respectively).
In order to obtain a formula for the supports $d_j(x,r)$ from \eqref{7} and \eqref{8} we still need to take into account the dependence on the prediction $r$.  
According to the mechanism proposed and presented previously, the value of $P(x_l^{(i)}|j)$ will decrease depending on the level of contamination $r_l$ of the channel (signal) $C_l$. This effect will be obtained for crisp and soft decision schemes by raising the appropriate probability to the power $r_l$.
This leads to the following final form of the classification functions of the naive Bayes algorithm $\psi$, and at the same time the entire two-stage patient's intention recognition system with identification of contaminated channels:

\begin{equation}   \label{9}
d_j(x,r)=c \cdot p_j \cdot \prod_{l=1}^{L} \left[ \prod_{i=1}^{d_l} P(x_l^{(i)}|j)\right]^{r_l} \  \  j \in \mathcal{M}.
\end{equation}
This formula defines an attribute-weighted Naive Bayes classifier~\cite{Kalra2022}

Note that by assuming conditional independence of features and defining decision functions as the product of probability distributions for single features, we do not need to train the classifier dynamically for each subsequent test object.
All that needs to be done is to determine all elementary probability distributions $P(x_l^{(i)}|j)$ (their number is $\sum_{l}d_l$) in a one-shot training process, which in this case consists of estimating these distributions from the training set treated as a random sample.
Finally, the functions \eqref{9} are dynamically determined and compared for different classes to identify the class with the highest support value.

\section{Experimental Setup}\label{sec:ExpSetup}

The experimental study is conducted to answer the research questions posed in Section~\ref{sec:Introduction}. To do so, we compared the following methods:
\begin{itemize*}
	\item \label{itm:methods:B}  \textbf{B}: A Naive Bayes classifier trained on the data coming from all available EMG channels.

	\item \label{itm:methods:EC} \textbf{EC}: Error-correcting output codes ensemble trained on the data from all available EMG channels~\cite{Sarabia2023}. We considered the following values of \textit{code\_size} parameter $\{2, 3, 4, 5, 6 \}$ The parameters are tuned with a simple search approach, using four-fold cross-validation.

	\item \label{itm:methods:NBH} \textbf{NBH}: Naive Bayes classifier with attributes selected in a crisp way.
	
	\item \label{itm:methods:NBS} \textbf{NBS}: Soft Naive Bayes approach. 

\end{itemize*}

In this study, we employed the One-class \textbf{SVM} classifier with RBF kernel~\cite{Fraser2014}. This is because in our previous studies this classifier achieved the best classification performance in the task of identifying noise-contaminated sEMG channels~\cite{Trajdos2025}. The classifier has a default method to adjust the parameter $\gamma$. The $\nu$ parameter of this classifier is tuned. The following values of $\nu$ are considered $\{ 0.1, 0.2, \ldots, 1.0\}$ 

To determine the best values for the parameters listed above, the following method is utilised. A four-fold, cross-validation approach is employed to create the training and validation sets. The validation segment is subsequently enhanced by incorporating artificial examples labelled as noise class. Artificial examples are generated using a uniform distribution in the classifier-specific input space $\mathcal{X}_l$. This is because the main motivation for using one-class classifiers instead of binary classifiers is to avoid making any assumptions about the outlier distribution. When the best value of the parameters is found, the final one-class classifier is trained using the entire training set. For parameter tuning, we use the balanced accuracy score as a quality measure.

The following versions of the Naive Bayes classifier are considered in the experiments:
\begin{itemize*}
  \item \label{itm:base:NBG} Naive Bayes classifier using a Gaussian distribution estimator (NBG).
  
  \item \label{itm:base:NBGMT} Naive Bayes classifier using a mixture of Gaussian distributions (NBGMT). The number of distributions is tuned using four-fold cross-validation and balanced accuracy as the quality criterion. The following numbers of Gaussian distributions are considered: $\{1, 3, 5 ,7 \}$. The seeds for Gaussian distributions are set using the K-means++ approach~\cite{Arthur2007}.
\end{itemize*}

We used the one-class and multiclass classifiers implemented in the scikit library~\cite{scikit-learn}. Since the soft outputs of the one-class classifiers used are incompatible with the model presented in~\eqref{9}, we applied the logistic regression model~\cite{Yu2010} to make them compatible with the model. Unless otherwise specified, the classifier parameters are set to their default values.

The signals used in these experiments come from the Web repository \footnote{\url{https://www.rami-khushaba.com/}}. The demographic information details for each amputee and the details of the dataset are presented in~\cite{AlTimemy2016}.
To unify the signals coming from different subjects, we used sEMG signals from the first 8 channels. We selected the signals associated with a low force level. In separating individual objects for particular classes, we used a non-overlapped segmentation scheme with a segment length of 500 ms. This resulted in 70 to 310 objects for each class.

To simulate real-world EMG and MMG signal contaminations, the following noise generation techniques are used~\cite{Farago2023,Boyer2023}
\begin{itemize*}

	\item Simulation of power grid whose frequency varies from 48 to 52 Hz. The amplitude of the inserted noise depends on the SNR.

	\item Signal attenuation that simulates the sensor losing contact with the skin. The attenuation level depends on the selected SNR.

	\item Gaussian noise. It simulates the general noises that may appear in the signal acquisition circuit.

	\item Simulation of non-linear amplifier characteristics for signals of high amplitude. This is done by non-linear clipping the peaks of the signal. The clipping level depends on the selected SNR. 

	\item Baseline wandering. This is simulated by injecting low-frequency (0.5 to 1.5 Hz) sinusoidal noise. The amplitude of the signal depends on the selected SNR.

\end{itemize*}

In experimental studies, we consider the following SNR levels $\{ 0,1,2, \ldots, 6, 10, 12 \}$.

The training and testing sets are obtained by a ten-fold cross-validation repeated 3 times. Each example from the testing dataset is then randomly contaminated with one of the above-mentioned noise types with a selected SNR level. The original testing set is finally extended using samples with artificial noise.

Feature vectors were created from raw EMG signals using the discrete wavelet transform technique. The \textit{db6} wavelet and three levels of decomposition were used. The following functions were calculated for the transformation coefficients \cite{MendesJunior2020}: MAV, SSC.

To asses the classification quality, we employed the following quality criteria:
\begin{itemize*}
  \item Balanced Accuracy (BAC),
  \item Cohen's kappa coefficient,
  \item Micro-averaged $F_1$ measure.
\end{itemize*}

The statistical significance of the results obtained was evaluated using the pairwise Wilcoxon signed rank test. Family-wise errors were using Holm's procedure. The significance level was set to $\alpha=0.05$~\cite{garcia2008extension}. For some analysis, the average rank approach is also used.
The experimental code is provided in~\footnote{\url{https://github.com/ptrajdos/CORES_2025.git}}

\section{Results and Discussion}\label{sec:ResAndDisc}

\begin{figure}[htb]
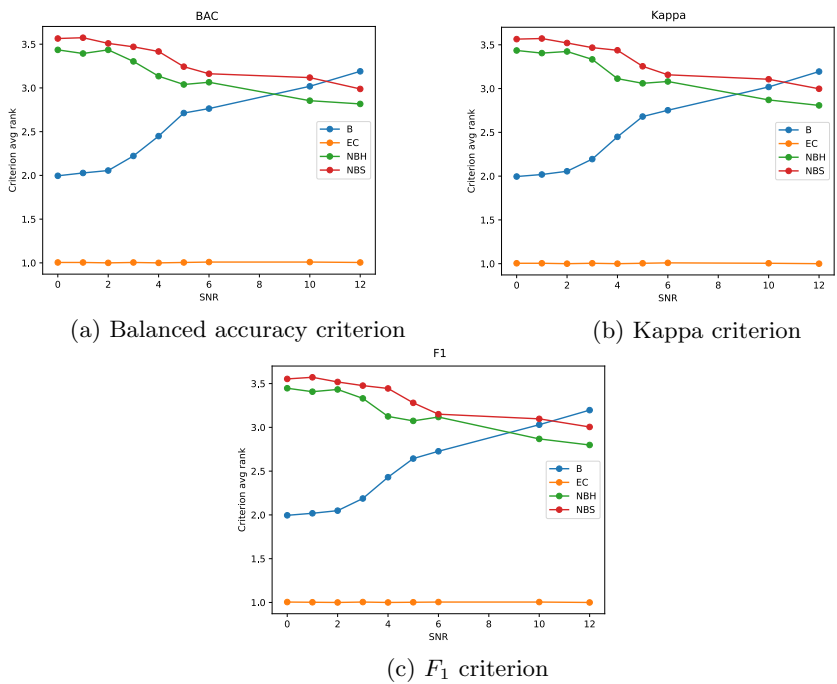

  \centering
  \begin{subfigure}[h]{.5\textwidth}
      \includegraphics[width=0.8\textwidth]{\ptFiguresDirectory{nbg_bac}}
  \caption{Balanced accuracy criterion}
  \label{figs:nbg_bac}
  \end{subfigure}%
  \begin{subfigure}[h]{.5\textwidth}
      \includegraphics[width=0.8\textwidth]{\ptFiguresDirectory{nbg_kappa}}
  \caption{Kappa criterion}
  \label{figs:nbg_kappa}
  \end{subfigure}
  \begin{subfigure}[h]{.5\textwidth}
      \includegraphics[width=0.8\textwidth]{\ptFiguresDirectory{nbg_f1}}
  \caption{$F_1$ criterion}
  \label{figs:nbg_f1}
  \end{subfigure}
  \caption{Classification quality for NBG classifier -- average ranks.}
  \label{figs:nbg_rnk}
\end{figure}

The average rank plots are presented in Figures~\ref{figs:nbg_rnk} and~\ref{figs:nbgmt_rnk}. The exact average rank values and the results of the statistical tests are shown in Tables~\ref{table:nbg_stats} and~\ref{table:nbgmt_stats}. Each table is divided into quality-criterion-specific sections. In each section, the column names are related to the investigated methods. In each row, the average ranks for a given SNR value are presented. The subscript under the average rank value presents the outcome of the statistical tests, with the subscript containing a comma-separated list of numbers or pause '--'. The list contains column numbers of methods that are significantly better than a given method. Pause means that the method is not significantly better than any other.

The results presented for both base classifiers \textbf{NBG} and \textbf{NBGMT} and all quality measures are pretty consistent. Let us describe the main thrends presented in the data. First of all, according to the average ranks, the \textbf{EC} ensemble is the worst method in this comparison. The results show that in the scenario considered, the decomposition based on the error-correcting output codes does not perform well in combination with the Naive Bayes classifier. The reason may be the fact that the method decomposes the multi-class classification problem ino multiple binary problems and then combines the results. The ensemble creation method is unaware that some data may be noisy.

The best performing methods for a wide range of SNR values are \textbf{NBS} and \textbf{NBH} methods. That is surely because the method works together with one-class-svm based outlier detectors. Consequently, they are aware of the presence of noise in the input data. The differences between them and the base Naive Bayes classifier (\textbf{B}) are greater when the SNR value is lower. That is, the dynamic Naive Bayes classifiers \textbf{NBH} and \textbf{NBS} tend to be better when the amount of noise introduced in the data is greater. This is because when the amount of noise is high, the base, Naive Bayes  classifier \textbf{B} is no longer robust against noise and elimination of affected channels causes the classification results to be better. When the amount of noise is lower (higher SNR values), the base Naive Bayes classifier seems to be robust to the noise present in the data and its results become comparable to \textbf{NBS} and \textbf{NBH} methods. This is especially visible in the case of weaker \textbf{NBG} base classifier. For \textbf{NBG} approach and SRN = 12, \textbf{B} method is significantly better than the dynamic approaches \textbf{NBH} and \textbf{NGS}. This may be due to the fact that, for SNR=12 the amount of noise is low enough for \textbf{B} to deal with it efficiently. In this scenario, any misclassification at the first level of the ensemble (one-class SVM classifier) has a significant negative impact on the overall classification accuracy. However, the phenomenon is not observed for the \textbf{NBGMT} approach.

When we compare the dynamic Naive Bayes approaches, some significant differences can also be observed. First, the \textbf{NBS} method tends to outperform the \textbf{NBH} method in most cases. In other cases the methods are comparable. There are no cases when \textbf{NBH} is significantly better than \textbf{NBS}. This means that weighting the attributes taking the amount of noise present in the related sEMG channel allows the classifier to perform better under noisy conditions. This is because the channel that contains a low amount of noise would give the method some information related to the object that is currently classified. The reason is related to the robustness of the base classifier to some degree of noise present in the data. Due to this robustness, completely eliminating the noisy channels is not a good strategy.

\vspace{-10pt}

\begin{figure}[htb]
  \centering
  \begin{subfigure}[h]{.5\textwidth}
      \includegraphics[width=0.8\textwidth]{\ptFiguresDirectory{nbgmt_bac}}
  \caption{Balanced accuracy criterion}
  \label{figs:nbgmt_bac}
  \end{subfigure}%
  \begin{subfigure}[h]{.5\textwidth}
      \includegraphics[width=0.8\textwidth]{\ptFiguresDirectory{nbgmt_kappa}}
  \caption{Kappa criterion}
  \label{figs:nbgmt_kappa}
  \end{subfigure}
  \begin{subfigure}[h]{.5\textwidth}
      \includegraphics[width=0.8\textwidth]{\ptFiguresDirectory{nbgmt_f1}}
  \caption{$F_1$ criterion}
  \label{figs:nbgmt_f1}
  \end{subfigure}
  \caption{Classification quality for NBGMT classifier -- average ranks.}
  \label{figs:nbgmt_rnk}
\end{figure}

{
\setlength\tabcolsep{2.0pt}%
\def\arraystretch{.8}%
\begin{table}[htb]
\centering
\scriptsize
\caption{Classification quality for NBG classifier, statistical tests.\label{table:nbg_stats}}
\begin{subtable}[h]{.33\textwidth}
  \centering
  \begin{tabular}{lllll}
    & \multicolumn{4}{c}{BAC} \\
    & B & EC & NBH & NBS \\
    \cmidrule(lr){2-5}
   SNR:12 & 3.190 & 1.005 & 2.817 & 2.988 \\
    & {\stiny 2,3,4} & -- & {\stiny 2} & {\stiny 2,3} \\
   SNR:10 & 3.019 & 1.009 & 2.854 & 3.118 \\
    & {\stiny 2} & -- & {\stiny 2} & {\stiny 2} \\
   SNR:6 & 2.764 & 1.009 & 3.065 & 3.162 \\
    & {\stiny 2} & -- & {\stiny 1,2} & {\stiny 1,2} \\
   SNR:5 & 2.713 & 1.005 & 3.039 & 3.243 \\
    & {\stiny 2} & -- & {\stiny 1,2} & {\stiny 1,2,3} \\
   SNR:4 & 2.449 & 1.000 & 3.134 & 3.417 \\
    & {\stiny 2} & -- & {\stiny 1,2} & {\stiny 1,2,3} \\
   SNR:3 & 2.222 & 1.005 & 3.303 & 3.470 \\
    & {\stiny 2} & -- & {\stiny 1,2} & {\stiny 1,2,3} \\
   SNR:2 & 2.056 & 1.000 & 3.435 & 3.509 \\
    & {\stiny 2} & -- & {\stiny 1,2} & {\stiny 1,2} \\
   SNR:1 & 2.028 & 1.005 & 3.394 & 3.574 \\
    & {\stiny 2} & -- & {\stiny 1,2} & {\stiny 1,2,3} \\
   SNR:0 & 1.995 & 1.005 & 3.435 & 3.565 \\
    & {\stiny 2} & -- & {\stiny 1,2} & {\stiny 1,2,3} \\
   \end{tabular}
\end{subtable}%
\begin{subtable}[h]{.33\textwidth}
  \centering
  \begin{tabular}{llll}
     \multicolumn{4}{c}{Kappa} \\
     B & EC & NBH & NBS \\
     \cmidrule(lr){1-4}
    3.194 & 1.000 & 2.808 & 2.998 \\
     {\stiny 2,3,4} & -- & {\stiny 2} & {\stiny 2,3} \\
    3.019 & 1.005 & 2.870 & 3.106 \\
    {\stiny 2} & -- & {\stiny 2} & {\stiny 2} \\
    2.752 & 1.009 & 3.081 & 3.157 \\
     {\stiny 2} & -- & {\stiny 1,2} & {\stiny 1,2} \\
   2.681 & 1.005 & 3.060 & 3.255 \\
    {\stiny 2} & -- & {\stiny 1,2} & {\stiny 1,2,3} \\
    2.449 & 1.000 & 3.113 & 3.438 \\
    {\stiny 2} & -- & {\stiny 1,2} & {\stiny 1,2,3} \\
    2.194 & 1.005 & 3.333 & 3.468 \\
    {\stiny 2} & -- & {\stiny 1,2} & {\stiny 1,2,3} \\
    2.056 & 1.000 & 3.424 & 3.521 \\
   {\stiny 2} & -- & {\stiny 1,2} & {\stiny 1,2} \\
   2.019 & 1.005 & 3.405 & 3.572 \\
   {\stiny 2} & -- & {\stiny 1,2} & {\stiny 1,2,3} \\
   1.995 & 1.005 & 3.435 & 3.565 \\
    {\stiny 2} & -- & {\stiny 1,2} & {\stiny 1,2,3} \\
   \end{tabular}
\end{subtable}%
\begin{subtable}[h]{.33\textwidth}
  \centering
  \begin{tabular}{llll}
\multicolumn{4}{c}{F1} \\
B & EC & NBH & NBS \\
\cmidrule(lr){1-4}
3.197 & 1.000 & 2.799 & 3.005 \\
{\stiny 2,3,4} & -- & {\stiny 2} & {\stiny 2,3} \\
3.030 & 1.005 & 2.868 & 3.097 \\
{\stiny 2} & -- & {\stiny 2} & {\stiny 2} \\
2.727 & 1.005 & 3.118 & 3.150 \\
{\stiny 2} & -- & {\stiny 1,2} & {\stiny 1,2} \\
2.644 & 1.002 & 3.074 & 3.280 \\
{\stiny 2} & -- & {\stiny 1,2} & {\stiny 1,2,3} \\
2.431 & 1.000 & 3.125 & 3.444 \\
{\stiny 2} & -- & {\stiny 1,2} & {\stiny 1,2,3} \\
2.188 & 1.005 & 3.331 & 3.477 \\
{\stiny 2} & -- & {\stiny 1,2} & {\stiny 1,2,3} \\
2.049 & 1.000 & 3.433 & 3.519 \\
{\stiny 2} & -- & {\stiny 1,2} & {\stiny 1,2} \\
2.019 & 1.002 & 3.407 & 3.572 \\
{\stiny 2} & -- & {\stiny 1,2} & {\stiny 1,2,3} \\
1.995 & 1.005 & 3.447 & 3.553 \\
{\stiny 2} & -- & {\stiny 1,2} & {\stiny 1,2,3} \\
   \end{tabular}
\end{subtable}
\end{table}
}

{
\setlength\tabcolsep{2.0pt}%
\def\arraystretch{.8}%
\begin{table}[htb]
\centering
\scriptsize
\caption{Classification quality for NBGMT classifier, statistical tests.\label{table:nbgmt_stats}}
\begin{subtable}[h]{.33\textwidth}
  \centering
  \begin{tabular}{lllll}
    & \multicolumn{4}{c}{BAC} \\
    & B & EC & NBH & NBS \\
    \cmidrule(lr){2-5}
   SNR:12 & 2.731 & 1.079 & 2.991 & 3.199 \\
    & {\stiny 2} & -- & {\stiny 1,2} & {\stiny 1,2,3} \\
   SNR:10 & 2.514 & 1.021 & 3.062 & 3.403 \\
    & {\stiny 2} & -- & {\stiny 1,2} & {\stiny 1,2,3} \\
   SNR:6 & 2.373 & 1.042 & 3.169 & 3.417 \\
    & {\stiny 2} & -- & {\stiny 1,2} & {\stiny 1,2,3} \\
   SNR:5 & 2.292 & 1.014 & 3.245 & 3.449 \\
    & {\stiny 2} & -- & {\stiny 1,2} & {\stiny 1,2,3} \\
   SNR:4 & 2.157 & 1.028 & 3.259 & 3.556 \\
    & {\stiny 2} & -- & {\stiny 1,2} & {\stiny 1,2,3} \\
   SNR:3 & 2.051 & 1.037 & 3.354 & 3.558 \\
    & {\stiny 2} & -- & {\stiny 1,2} & {\stiny 1,2,3} \\
   SNR:2 & 1.968 & 1.065 & 3.389 & 3.579 \\
    & {\stiny 2} & -- & {\stiny 1,2} & {\stiny 1,2,3} \\
   SNR:1 & 1.926 & 1.074 & 3.438 & 3.562 \\
    & {\stiny 2} & -- & {\stiny 1,2} & {\stiny 1,2,3} \\
   SNR:0 & 1.954 & 1.051 & 3.463 & 3.532 \\
    & {\stiny 2} & -- & {\stiny 1,2} & {\stiny 1,2} \\
   \end{tabular}
\end{subtable}%
\begin{subtable}[h]{.33\textwidth}
  \centering
  \begin{tabular}{lllll}
\multicolumn{4}{c}{Kappa} \\
B & EC & NBH & NBS \\
\cmidrule(lr){1-4}
2.725 & 1.083 & 3.007 & 3.185 \\
{\stiny 2} & -- & {\stiny 1,2} & {\stiny 1,2,3} \\
2.502 & 1.028 & 3.067 & 3.403 \\
{\stiny 2} & -- & {\stiny 1,2} & {\stiny 1,2,3} \\
2.343 & 1.051 & 3.178 & 3.428 \\
{\stiny 2} & -- & {\stiny 1,2} & {\stiny 1,2,3} \\
2.280 & 1.019 & 3.257 & 3.444 \\
{\stiny 2} & -- & {\stiny 1,2} & {\stiny 1,2,3} \\
2.125 & 1.028 & 3.285 & 3.562 \\
{\stiny 2} & -- & {\stiny 1,2} & {\stiny 1,2,3} \\
2.037 & 1.037 & 3.361 & 3.565 \\
{\stiny 2} & -- & {\stiny 1,2} & {\stiny 1,2,3} \\
1.968 & 1.065 & 3.398 & 3.569 \\
{\stiny 2} & -- & {\stiny 1,2} & {\stiny 1,2,3} \\
1.912 & 1.088 & 3.412 & 3.588 \\
{\stiny 2} & -- & {\stiny 1,2} & {\stiny 1,2} \\
1.954 & 1.046 & 3.463 & 3.537 \\
{\stiny 2} & -- & {\stiny 1,2} & {\stiny 1,2} \\
   \end{tabular}
\end{subtable}%
\begin{subtable}[h]{.33\textwidth}
  \centering
  \begin{tabular}{lllll}
    \multicolumn{4}{c}{F1} \\
B & EC & NBH & NBS \\
\cmidrule(lr){1-4}
2.720 & 1.076 & 2.998 & 3.206 \\
{\stiny 2} & -- & {\stiny 1,2} & {\stiny 1,2,3} \\
2.493 & 1.025 & 3.086 & 3.396 \\
{\stiny 2} & -- & {\stiny 1,2} & {\stiny 1,2,3} \\
2.324 & 1.046 & 3.194 & 3.435 \\
{\stiny 2} & -- & {\stiny 1,2} & {\stiny 1,2,3} \\
2.264 & 1.019 & 3.257 & 3.461 \\
{\stiny 2} & -- & {\stiny 1,2} & {\stiny 1,2,3} \\
2.134 & 1.025 & 3.278 & 3.562 \\
{\stiny 2} & -- & {\stiny 1,2} & {\stiny 1,2,3} \\
2.030 & 1.035 & 3.373 & 3.562 \\
{\stiny 2} & -- & {\stiny 1,2} & {\stiny 1,2,3} \\
1.965 & 1.065 & 3.410 & 3.560 \\
{\stiny 2} & -- & {\stiny 1,2} & {\stiny 1,2,3} \\
1.917 & 1.083 & 3.435 & 3.565 \\
{\stiny 2} & -- & {\stiny 1,2} & {\stiny 1,2} \\
1.956 & 1.044 & 3.465 & 3.535 \\
{\stiny 2} & -- & {\stiny 1,2} & {\stiny 1,2} \\
   \end{tabular}
\end{subtable}
\end{table}
}

\section{Conclusions}\label{sec:Conclusions}

The objective of this paper is to develop a multiclass recognition system capable of identifying signal contamination and leveraging these contamination-related data to improve the classification of contaminated samples. The proposed method is based on two cooperating classification systems in a cascade structure. The first system is an ensemble of one-class classifiers whose task is to recognise contaminated sEMG signals. The second system is a Naive Bayes classifier (NB) working in dynamic mode and using the information provided by the previous one-class ensemble system. This is done either by eliminating features originating from contaminated signals or by weighting the influence of features depending on the degree of channel contamination (for the soft scheme). The training and validation sets are subsequently enhanced by incorporating artificial examples labelled as noise class. The greatest advantage of the dynamic Naive Bayes classifier is that it does not need any retraining to order to change classification model by removing or changing the weights of attributes. In addition, the experimental study shows that using the information from the one-class classifiers allows the entire ensemble to improve the classification quality significantly compared to the reference methods. The experiments also show that the sof version of the dynamic Naive Bayec classifier performs significantly better than the crisp version. The results are very promising, and the concept of using a dynamic Naive Bayes classifier in the context of classification of contaminated sEMG signals should be further explored.


\bibliography{bibliography}

\end{document}